\documentclass[twocolumn,american]{revtex4-2}
\usepackage[T1]{fontenc}
\usepackage[latin9]{inputenc}
\usepackage{refstyle}
\usepackage{bm}
\usepackage{amsbsy}
\usepackage{amstext}
\usepackage{graphicx}
\usepackage{esint}



\begin{document}
	\title{Magnetically confined charged particles: From steep density profiles
		to the breaking of the adiabatic invariant}
	\author{Aur\'elien Cordonnier}
	\affiliation{Aix Marseille Univ, Universit\'e de Toulon, CNRS, CPT, Marseille, France}
	\author{Yohann Lebouazda}
	\affiliation{Aix Marseille Univ, Universit\'e de Toulon, CNRS, CPT, Marseille, France}
	\author{Xavier Leoncini}
	\affiliation{Aix Marseille Univ, Universit\'e de Toulon, CNRS, CPT, Marseille, France}
	\author{Guilhem Dif-Pradalier}
	\affiliation{CEA, IRFM, F-13108 St. Paul-lez-Durance cedex, France}
	\begin{abstract}
		This study examines the stability of Vlasov equilibrium solutions for magnetically confined plasmas, derived through the principle of maximum entropy. By treating the toroidal limit as a perturbation from an analytical cylindrical solution, we demonstrate that these equilibria align well with the inviscid magnetohydrodynamic (MHD) description. Using the aspect ratio as a perturbation parameter, we compute particle trajectories sampled from the kinetic equilibrium distribution, confirming the overall stability of the solutions.
		However, under burning plasma conditions, chaotic dynamics emerge for particles with supra-thermal and even thermal energies. This destroys the adiabatic invariance of the magnetic moment. The exact consequences are unclear, but they could undermine the foundational assumptions of gyrokinetic modelling in burning plasmas. Nevertheless, these results suggest the possibility of unaccounted transport losses in future burning plasma operations. The interplay between turbulence and energetic particles in the presence of Hamiltonian chaos certainly warrants further investigation. 
	\end{abstract}
	\maketitle
	In order to confine hot plasmas and achieve fusion, the main idea is to rely on a strong magnetic field. One of the devices of choice for this purpose is the tokamak. In this configuration, the poloidal part of the confining magnetic field is generated by the plasma current itself. It is also widely accepted that the plasma distribution profile corresponds to an out-of-equilibrium configuration with strong density and temperature gradients. At a global scale, equilibria computed in the framework of magnetohydrodynamics (MHD) are usually considered to be solutions of the so-called Grad--Shafranov equation \cite{grad_1958,Shafravov66}.
	
	Recent work has shown that such non-equilibrium plasmas may, in fact, remain close to thermodynamic equilibrium. In particular, in \cite{cordonnier_full_2022}, the authors used the maximum entropy principle to derive self-consistent Vlasov--Maxwell equilibrium profiles for a tokamak approximated as an infinite-aspect-ratio cylinder. These maximum entropy equilibria feature strongly shaped density and temperature profiles (in the sense of kinetic temperature) and are peculiar in that they introduce two additional thermodynamic parameters (appearing as Lagrange multipliers) associated with the toroidal and poloidal momentum fluxes. These parameters are related to, but not the same as, the plasma current, which is the central quantity in the Grad--Shafranov framework. Because plasma momentum is indirectly linked to the plasma current, the Grad--Shafranov and maximum entropy equilibria lead to subtle differences in the kinetic profiles and magnetic shear computed by both approaches. These differences affect how flows and flow shear develop nonlinearly. Subtle differences in the background equilibrium can thus lead to substantial discrepancies in the emergence and maintenance of transport barriers or transport bifurcations \cite{connor_review_2004,wolf_internal_2003, firpo_interplay_2024}.
	
	Another important topic concerns variations in the magnetic moment $\mu$, which are linked to the motion of charged particles in magnetic fields \cite{dragt_insolubility_1976,weitzner_nonperiodicity_1999}. Assuming a large background guiding field, as in fusion, $\mu$ is an adiabatic invariant. Most fundamental approaches to transport in fusion plasmas rely on gyrokinetic theory \cite{brizard_foundations_2007}, which is constructed such that $\mu$ is an invariant at any given order. It is crucial to test whether classes of particles or locations in the plasma volume break this adiabatic invariant for a large number of trajectories in settings that are relevant for fusion plasmas. 
	
	In this letter, we address both of the above problems. First, we assess the stability of the maximum entropy solutions, showing that they coincide with the classical MHD equilibria in the zero-viscosity limit. We then examine the stability of these thermodynamic equilibria when transitioning from a cylinder to a torus \cite{laribi_influence_2019,ogawa_study_2016,ogawa_tailoring_2019,ogawa_full_2016}. Finally, following  \cite{cambon_chaotic_2014}, we analyse the stability of trajectories and quantify the conservation of magnetic moment for various classes of particles. Many energetic particle trajectories exhibit chaotic behaviour associated with a violation of the adiabatic invariant, which can pose a challenge when modelling burning plasmas. 
	
\begin{figure}
		\centering{}
		\includegraphics[width=4cm]{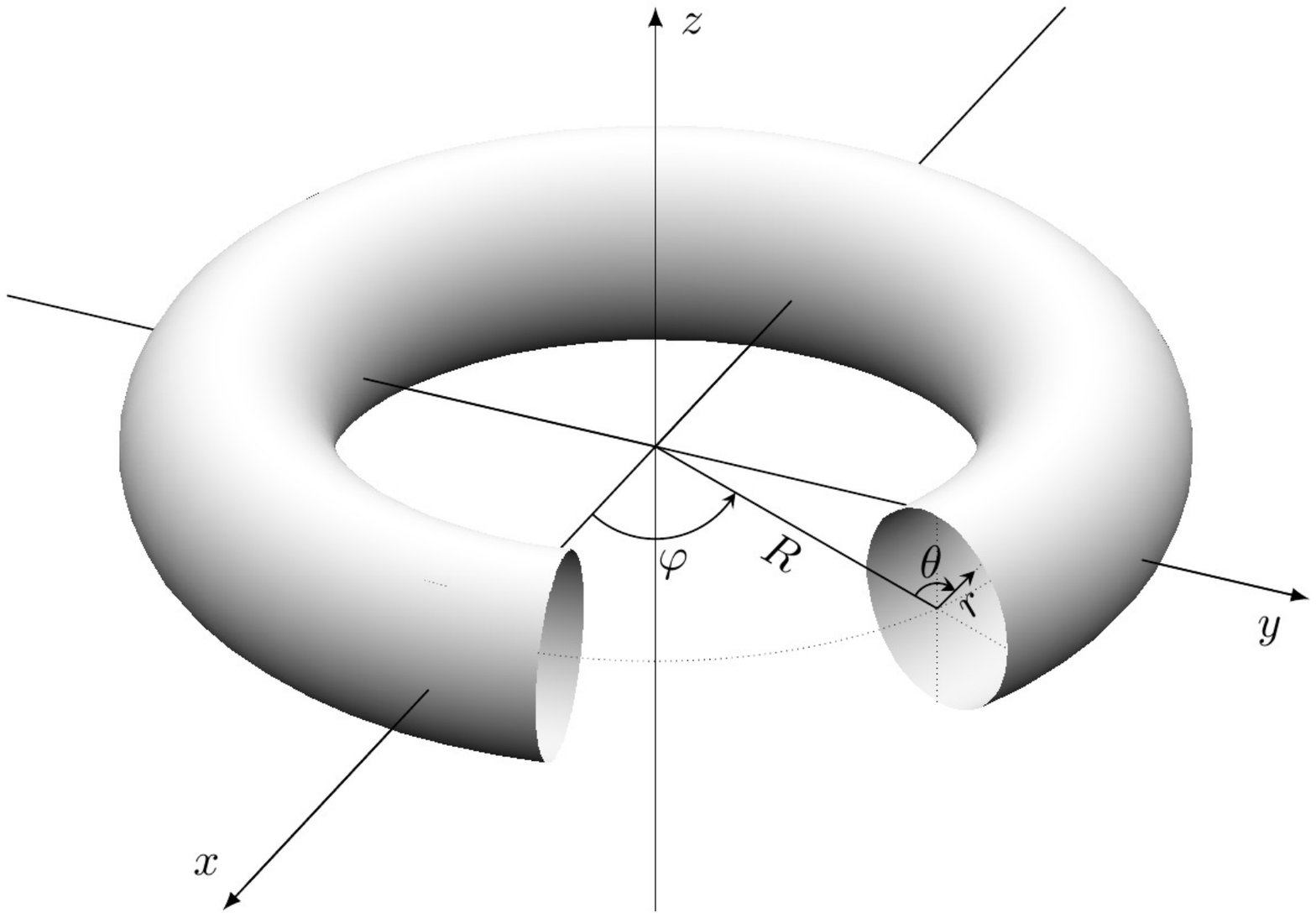}
		\includegraphics[width=4cm]{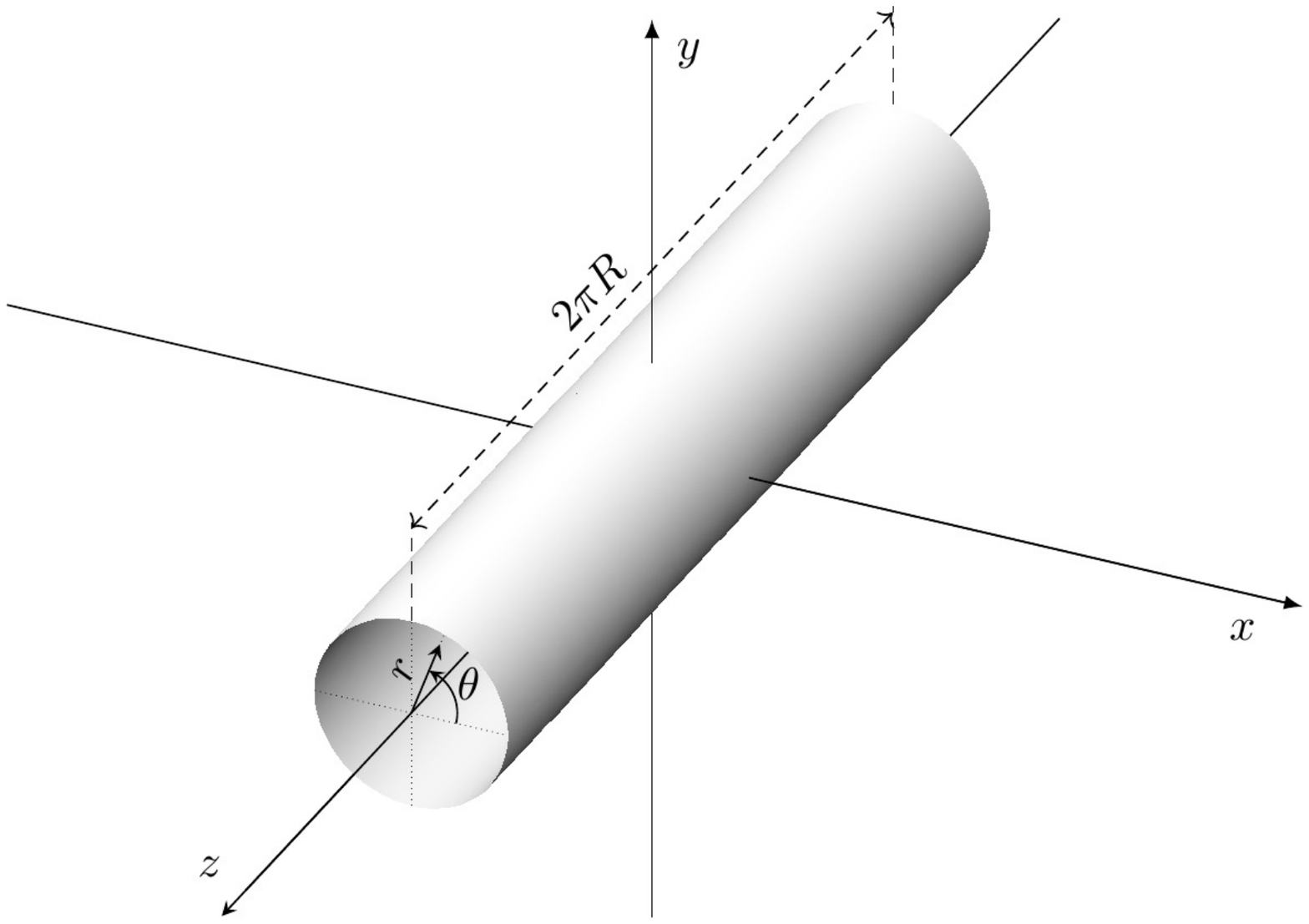}
		\caption{Notations used in the paper. In the infinite aspect ratio limit ($R\rightarrow\infty$),
			the torus locally becomes a cylinder, the angle $\varphi$ matching
			$\frac{z}{R}$. A rotational invariance around the $\theta$ direction is recovered,
			and the rotational invariance around $\varphi$ becomes a translational
			one along $z$ with $2\pi R$ periodic boundary conditions}.\label{fig:Geometry}
	\end{figure}

	Fully self-consistent solutions in cylindrical geometry have been proposed in \cite{cordonnier_full_2022}. Starting from the same point, we consider the magnetic field of an ideal toroidal tokamak, taking an infinite aspect ratio limit to yield a cylinder of the form:
	\begin{equation}
		\mathbf{B}(r)=B_{0}\left[g(r)\,\text{\ensuremath{\mathbf{e}}}_{\theta}+\left(1+k(r)\right)\,\text{\ensuremath{\mathbf{e}}}_{z}\right],\label{eq:B_field}
	\end{equation}
	where $\mathbf{e}_{\theta}$ and $\mathbf{e}_{z}$ are the unit vectors in the poloidal and axial directions, respectively, and $r$ denotes the radial direction in cylindrical coordinates (see Fig.\ref{fig:Geometry}). In this setting, it was shown that separatrices can appear in the phase space of the integrable motion of charged particles, near the center of the distribution, even for thermal particles. In the context of magnetic fusion, this would correspond to particles with an energy of about $10~\mathrm{keV}$. 
	In Coulomb gauge, Eq.(\ref{eq:B_field}) is related to the magnetic vector potential:
	\begin{equation}
		\mathbf{A}(r)=B_{0}\left[\left(\frac{r}{2}+\frac{K(r)}{r}\right)\,\mathbf{e}_{\theta}-G(r)\,\mathbf{e}_{z}\right]\:,
		\label{eq:A_field}
	\end{equation}
	where $G(r)=\intop_{0}^{r}g(u)\,\mathrm{d}u\:$, and $K(r)=\intop_{0}^{r}uk(u)\,\mathrm{d}u$.
	In this approximation of a torus as an infinite-aspect-ratio cylinder, particle motion is integrable, and the Hamiltonian with unit mass and charge and vanishing electric field is given by:
	\begin{equation}
		H=\frac{1}{2}\left[p_{r}^{2}+\left(\frac{p_{\theta}}{r}-B_{0}\left[\frac{r}{2}+\frac{K}{r}\right]\right)^{2}+\left(p_{z}+B_{0}G\right)^{2}\right]\:,
		\label{eq:H}
	\end{equation}
	where $p_{r}$, $p_{\theta}$ and $p_{z}$ represent the cylindrical components of the canonical momentum and are related to the continuous symmetries of the problem. Thus, the angular momentum $p_{\theta}$, the linear momentum $p_{z}$ and the kinetic energy (\ref{eq:H}) are the three constants of the particle motion ($\frac{p_{\theta}}{r}$ is related to the vector momentum that is not conserved).
	\smallskip
	
	{\noindent\it The maximum entropy approach.} We can then construct stationary one-particle density functions,  $f(r,\theta,z,p_{r},p_{\theta},p_{z})$, by selecting distributions that commute with the Hamiltonian (Eq.(\ref{eq:H})) and maximise the Boltzmann--Gibbs--Shannon entropy, subject to constraints related to the conservation of energy, momentum $p_{\theta}$ and $p_{z}$, and the number of particles (see \cite{ogawa_tailoring_2019,cordonnier_full_2022,laribi_influence_2019} for details). The associated Lagrange multipliers are denoted by  $\beta$, $\gamma_{\theta}$, $\gamma_{z}$ and $\mu_{N}$, and the resulting distributions are of the form:
	\begin{equation}
		f\sim e^{-\beta H-\gamma_{z}p_{z}-\gamma_{\theta}p_{\theta}-\mu_{N}}\:,
		\label{eq:Expression_f}
	\end{equation}
	up to a normalization (fixed by $\mu_{N}$). From (\ref{eq:Expression_f}), we can derive equilibrium cylindrical density profiles, $\rho$, and currents that are functions of $r$ only:
	\begin{equation}
		\rho(r)=\rho_{0} \mathrm{e}^{-ar^{2}-bG(r)-cK(r)}\:,
		\label{eq:DensCyl}
	\end{equation} 
	with $a=\frac{\gamma_{\theta}}{2}\left(B_{0}-\frac{\gamma_{\theta}}{\beta}\right)$, $b=-B_{0}\gamma_{z}$, $c=B_{0}\gamma_{\theta}$, three constants related to the average flow velocities and temperature, and $\rho_{0}=\frac{N}{\mathcal{V}}$, the density at the center.
	Linking back the current to the vector potential, we end up with a second-order ODE to solve for the current density $j$:
	\begin{equation}
		\frac{\mathrm{d}^{2}j}{\mathrm{d}r^{2}}=\left(\frac{1}{j}\frac{\mathrm{d}j}{\mathrm{d}r}-\frac{1-r^{2}}{r\left(1+r^{2}\right)}\right)\frac{\mathrm{d}j}{\mathrm{d}r}-\left(\frac{2\alpha_{0}}{1+r^{2}}+(1+r^{2})j\right)j\:,
		\label{eq:SelfCurrentTilde}
	\end{equation}
	where $j$ is proportional to the current density $j_{z}=-\frac{\gamma_{z}}{\beta}\rho$ along $z$, and $\alpha_0$ is a constant of integration (see \cite{cordonnier_full_2022} for details). Solving this closing equation (\ref{eq:SelfCurrentTilde}) leads to non-trivial plasma profiles at thermodynamic equilibrium. Before investigating their stability, let us first consider the twin problem from a magnetohydrodynamic (MHD) viewpoint. 
	\smallskip
	
	{\noindent\it The MHD approach.} We consider the same system, at equilibrium ($\partial/\partial t=0$), and  with the same symmetries: all quantities are functions of the radius $r$ only. The plasma is described by an ideal gas equation of state $P(r)=\rho(r)/\beta$. Here, $P$ is the pressure, $\rho$ the mass density, $\mathbf{u}$ the plasma fluid velocity and, with unit charge, the current reads $\mathbf{j}(\mathit{r})=\rho(\mathit{r})\mathbf{u}(\mathit{r})$. Mass conservation implies $u_{r}=0$, and $\boldsymbol{\nabla}\cdot\mathbf{B}=0$ implies  $B_{r}=0$. Substituting these into the Navier--Stokes equation with non-zero viscosity $\nu$:
	\begin{equation}
		\rho(\mathbf{u}\cdot\boldsymbol{\nabla})\mathbf{u}=-\boldsymbol{\nabla} P+\mathbf{j}\times\mathbf{B}+\rho\nu{\boldsymbol{\nabla}}^2\mathbf{u}\:,
		\label{eq:Navier-Stokes}
	\end{equation}
	leads to ${\boldsymbol{\nabla}}^2\mathbf{u}=\mathbf{0}$, which implies $u_{\theta}=\text{\ensuremath{\Omega r}}$ and $u_{z}=U$, where $\Omega$ and $U$ are constants. Projecting onto the radial direction now yields an equation for the density:
	\begin{equation}
		\frac{\mathrm{d}\rho}{\mathrm{d}r}+\beta\left(B_{\theta}U-B_{z}r\Omega-r\Omega^{2}\right)\rho=0\:,
		\label{eq:N-S_sur_rho}
	\end{equation}
	and the dependence of $\mathbf{B}$ on $\rho$ follows from the Maxwell--Amp\`ere equations:
	\begin{equation}
		\left\{
		\begin{array}{cc}
			\displaystyle -\frac{\mathrm{d}B_{z}}{\mathrm{d}r} &= \mu_{0}j_{\theta} = \mu_{0}\Omega r\rho \:, \label{eq:Max-Ampa} \\ 
			\\
			\displaystyle \frac{\mathrm{d}(rB_{\theta})}{\mathrm{d}r} &= \mu_{0}rj_{z} = \mu_{0}Ur\rho \:, \label{eq:Max-Ampb}
		\end{array}
		\right.
	\end{equation}
	from which
	\begin{equation}
		\frac{\mathrm{d}(rB_{\theta})}{\mathrm{d}r}+\frac{U}{\Omega}\frac{\mathrm{d}B_{z}}{\mathrm{d}r}=0\ \Rightarrow\ rB_{\theta}+\frac{U}{\Omega}B_{z}=\alpha_{0} \:,
		\label{eq:Lien_B_z_et_B_theta}
	\end{equation}
	follows, where $\alpha_{0}$ is a constant of integration. After deriving Eq.(\ref{eq:N-S_sur_rho}) and combining it with Eqs. (\ref{eq:Max-Ampa}) and (\ref{eq:Lien_B_z_et_B_theta}), we recover an equation formally identical to Eq.~(\ref{eq:SelfCurrentTilde}). This equivalence demonstrates that maximizing the entropy yields the stationary equilibrium distribution of ideal magnetohydrodynamics (MHD) in the inviscid limit, due to the conservative nature of the Vlasov equation. This is reminiscent of what happens for the Euler equation, among the infinite number of stationary solutions, those that maximize entropy actually select the solutions to Navier-Stokes' equations when viscosity tends toward zero (see \cite{bardos_vanishing_2012,chen_principle_2024} for further details).
	Consequently, these solutions are expected to exhibit stability, at least with respect to microscopic perturbations.
	\smallskip
	
	{\noindent\it Stability of the solutions.} These distributions have been obtained in the cylindrical limit. We now assess their stability in the presence of curvature using a simplified, perturbative approach, whereby we observe how an ensemble of charged particles picked from the cylindrical equilibrium distribution evolve in a toroidal tokamak-like setting.
	
	\begin{figure}
		\begin{centering}
			\includegraphics[width=7.2cm]{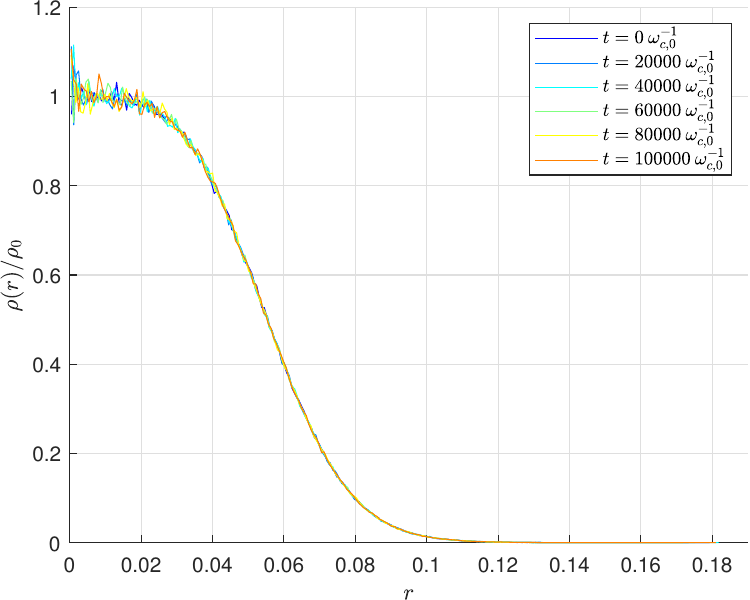}
		\end{centering}
		\caption{Stability of radial density in toroidal geometry for $R=100~\mathrm{m}$, $1/\beta=10~\mathrm{keV}$, 
			$B_{0}=1~\mathrm{T}$, $a=430$ and $b=-13$. The simulation was performed with $2^{20}$ particules, with a time step $\Delta t=0.025~\omega_{c,0}^{-1}$.\label{fig:Stability}}
	\end{figure}
	In order to mimic the self-consistent equilibria, we tailor a close Hamiltonian system, in the spirit of \cite{ogawa_tailoring_2019}, by picking the function $G$ among the eigenfunctions of the Laplace operator on the magnetic axis, and by setting $K=0$. 
	In this configuration, we obtain --- non-self-consistent --- equilibrium distributions for passive particles evolving in an ambient bulk field (without retroactive effect on it). Note that maximizing entropy implies that these distributions are of the form (\ref{eq:Expression_f}) and lead to densities of the form (\ref{eq:DensCyl}). Thus, considering $G(r)=G_{0}J_{2}(\lambda r)$, where $J_{2}$ is the third Bessel function of first kind, $G_{0}$ and $\lambda$ are two fitting parameters, we specify the magnetic field such that the distribution exhibits steep density profiles and, in particular, comparable phase space structures. With these settings, following \cite{cambon_chaotic_2014}, we can easily trace back to a compatible toroidal Hamiltonian system that allows us to recover the asymptotic Hamiltonian system. Therefore, the magnetic field in toroidal geometry reads:
	\begin{equation}
		\mathbf{B}(r,\theta)=\frac{B_{0}R}{\xi(r,\theta)}\left[g(r)\:\hat{\mathbf{e}}_{\theta}+\hat{\mathbf{e}}_{\varphi}\right]\:,
		\label{eq:champ_B}
	\end{equation} 
	which corresponds to a vector potential (in Coulomb gauge):
	\begin{equation}
		\mathbf{A}(r,\theta)=B_{0}R\ln\left(\frac{R}{\xi(r,\theta)}\right)\,\hat{\mathbf{e}}_{z}-\frac{B_{0}R}{\xi(r,\theta)}G(r)\,\hat{\mathbf{e}}_{\varphi}\:,\label{eq:Potentiel_A}
	\end{equation}
	where $\xi(r,\theta)=R-r\cos\theta$ is the distance to the axis of the torus, $R$ is the major radius and $\hat{\mathbf{e}}_{z}$ the unit vector along the axis of symmetry of the torus. By setting $G_{0}=\frac{1}{10}$ and $\lambda=10\lambda_{1}$ with $\lambda_{1}$ the first zero of $J_{2}$, we obtain distributions that show a bifurcation in density, around  $a=430$ and $b=-13$, between ``centered'' and ``off-centered'' profiles, as in the self-consistent case (see \cite{cordonnier_full_2022}). These values imply average flow velocities $\left\langle v_{\theta}\right\rangle \sim\left\langle v_{\varphi}\right\rangle \sim10^{5}~\mathrm{m}\thinspace\mathrm{s}^{-1}$, as well as steep current and density profiles whose tails are located at approximately $r\sim0.1~\mathrm{m}$, assuming $R$ in meters. We simulate equilibria using about $10^{6}$ particles over timescales of the order of $10^{-3}~\mathrm{s}$ for protons in a magnetic field of the order of $1~\mathrm{T}$, at typical temperatures of the order of $10~\mathrm{keV}$. The particle trajectories are computed individually using a sixth-order Gauss--Legendre symplectic scheme with a typical time step of $\frac{1}{40}$ (one correspond to a typical cyclotron period $\omega_{c,0}^{-1}=\frac{m}{eB_{0}}\sim10^{-8}~\mathrm{s}$). It should be noted that without self-consistency, the parameter $N$ --- which plays no role in the simulation --- is not constrained once $\gamma_{\theta}$, $\gamma_{z}$ and $\beta$ are fixed. Nevertheless, the self-consistent model predicts a density at the center of the order of $10^{20}~\mathrm{m^{-3}}$ for the range of parameters involved. The results are displayed in \figref{Stability}. Remarkably, the profiles remain stable over time, even when the aspect ratio is small, which challenges the validity of the perturbative analysis. It is worth noting that we observed the same stability for each case studied, which was unexpected, especially for small aspect ratios. In summary, the analogy with the MHD equilibrium and the stability of the cylindrical distribution in a toroidal setting are good indications of the relative stability of the full self-consistent distributions discussed in Cordonnier et al. \cite{cordonnier_full_2022}.
	\begin{figure}
		\begin{centering}
			\includegraphics[width=6.8cm]{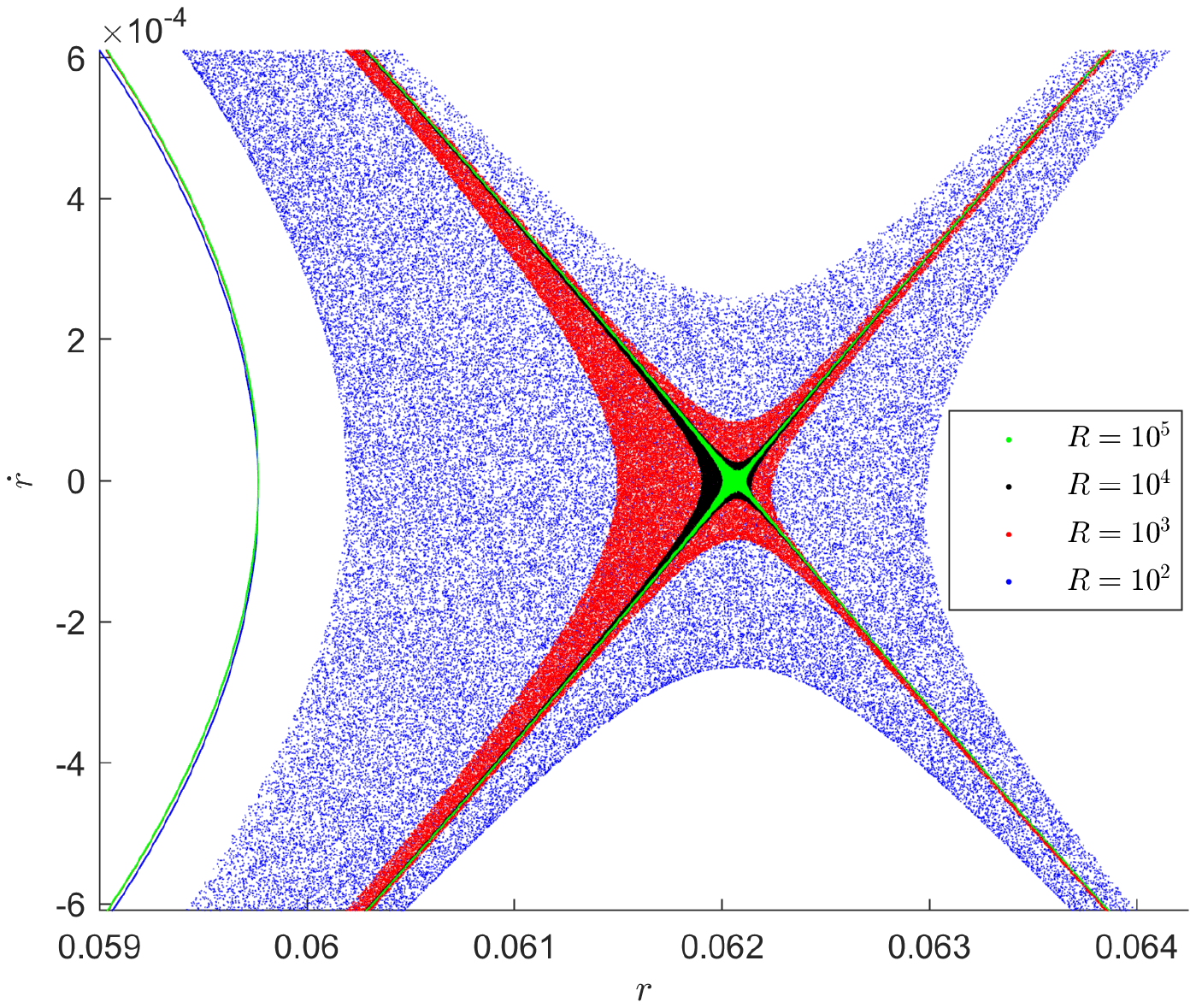}
			\includegraphics[width=6.8cm]{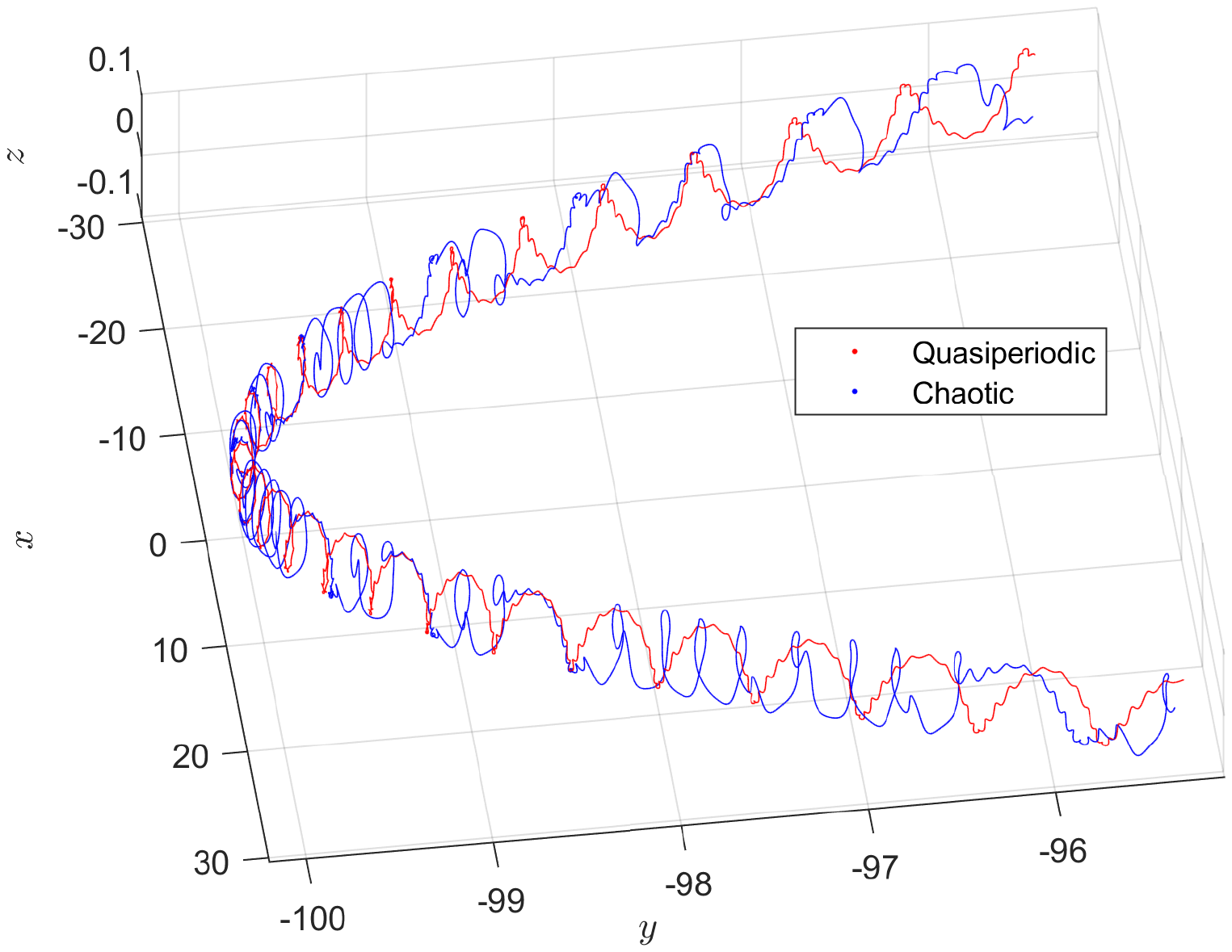}
		\end{centering}
		\caption{Dynamics of two neighboring particles obtained from $p_{\theta}\simeq0.00175$ and $p_{z}\simeq-0.03453$.
			Top figure: Poincar\'e sections, for $R=10^{5},10^{4},10^{3},10^{2}$.
			On the right part of the plot we see the evolution of  the stochastic layer for a particle with an energy $E\simeq10.00~\mathrm{keV}$. On the left is the evolution of a quasiperiodic trajectory without chaos for a lower energy, $E\simeq9.97~\mathrm{keV}$.
			The trajectories are followed for  $t=4\times10^{7}\,\omega_{c,0}^{-1}$, with $\Delta t=0.01~\omega_{c,0}^{-1}$.
			Bottom figure: Three-dimensional plots of part of the same trajectories in dashed red (quasiperiodic), in dashed blue (chaotic), for $R = 100$. \label{fig:-evolution-of-section}}
	\end{figure}
	\smallskip
	
	{\noindent\it Onset of chaos for thermal particles, adiabaticity of $\mu$ violated.}
	We now turn our attention to the motion of individual particles within a toroidal geometry. In a cylindrical setting, particle trajectories  are completely integrable, and depending on the field--particle configuration at equilibrium, a set of X-points --- hyperbolic fixed points --- may be present, in the $(r,p_{r})$ planes \cite{cordonnier_full_2022}. However, when transitioning to a torus, the rotational symmetry about the cylindrical axis is broken, eliminating one constant of motion, $p_{\theta}$, leaving only two: the angular momentum $p_{\varphi}$ associated with the toroidal symmetry, and the total energy. Since particle motion involves three degrees of freedom but retains only two constants of motion, the system becomes non-integrable in the Liouville--Arnold sense, giving rise to the possibility of Hamiltonian chaos. It is worth noting that the magnetic field lines themselves remain fully integrable, winding smoothly around nested tori. Yet, the quasiperiodic dynamics in the vicinity of these potential hyperbolic points is likely to be destroyed by perturbations, particularly by curvature.
	To investigate the onset of chaos, we focus on the neighborhood of these critical points and construct Poincar\'e sections in the $(r,\dot{r})$ planes ($p_{r}=\dot{r}$ from a cylindrical perspective), at a fixed ``cylindrical energy'', using the three action constants derived from the integrable cylindrical geometry  \cite{cambon_chaotic_2014}. Poincar\'e sections for particles with an energy of approximately $10~\mathrm{keV}$ --- a value typical of thermal particles in burning fusion plasmas --- was computed in the simplified magnetic configuration described above, near an aforementioned X-point. The results, presented in \figref{-evolution-of-section}, reveal Hamiltonian chaos, characterised by a growing stochastic layer as the aspect ratio decreases with $R$. For $R=10$ (not shown in \figref{-evolution-of-section}), these phase space structures encompass the two trajectories, which ultimately prove to be chaotic. It should be noted that the field configuration implies the presence of a set of thermal-energy hyperbolic fixed points, between approximately $r=0.04$ and $r=0.08$, which is more or less covered by the distribution, depending on $a$ and $b$ (with $B_{0}\sim1~\mathrm{T}$ and $\beta^{-1}\sim10~\mathrm{keV}$). For distributions such as the one illustrated in \figref{Stability}, roughly $15\%$ of particles may have chaotic trajectories as $R$ decreases.
	\begin{figure}
		\begin{centering}
			\includegraphics[width=6.8cm]{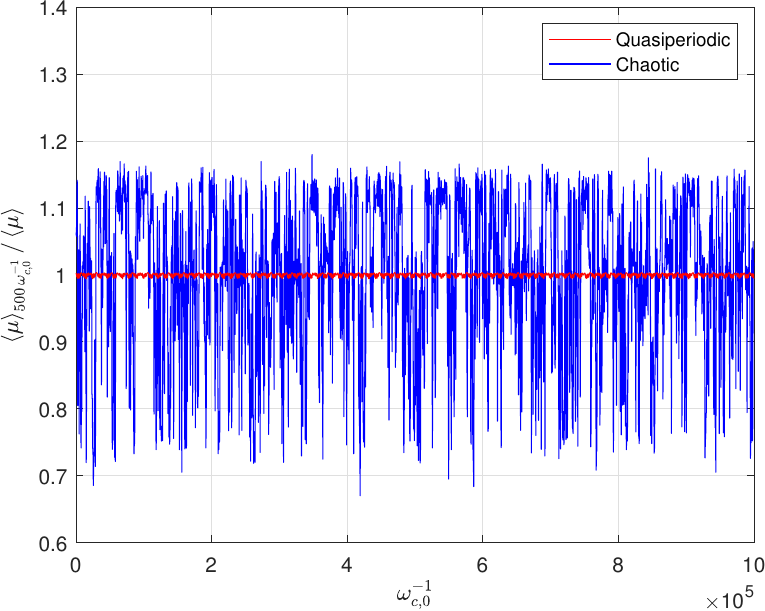}
			\includegraphics[width=7.1cm]{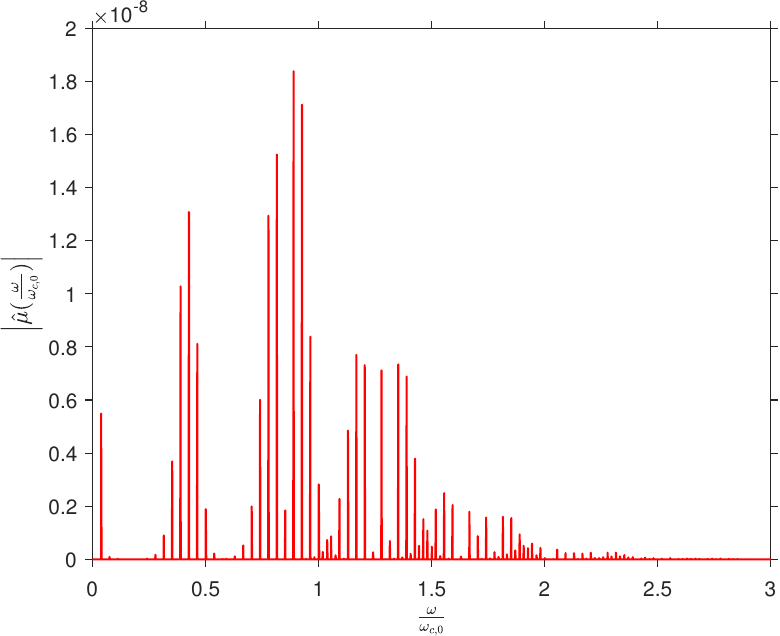}
			\includegraphics[width=7.1cm]{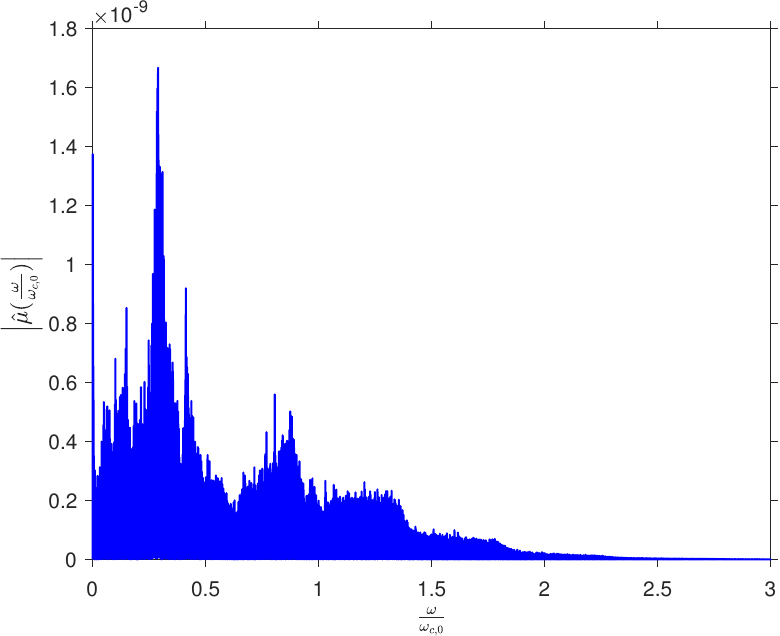}
		\end{centering}
		\caption{Magnetic moment for $R=100$ of the two neighboring trajectories depicted in \figref{-evolution-of-section} (from $p_{\theta}\simeq0.00175$ and $p_{z}\simeq-0.03453$), and computed with $\Delta t=0.01~\omega_{c,0}^{-1}$. Top figure: Relative fluctuations in the magnetic momentum averaged over a finite time period of $500\,\omega_{c,0}^{-1}$. Central panel: Fourier spectrum of $\mu$ for the quasiperiodic trajectory (at $E\simeq9.97~\mathrm{keV}$). Bottom panel: Fourier spectrum of $\mu$ for the chaotic trajectory (at $E\simeq10.00~\mathrm{keV}$).}
		\label{fig:Mu_mean}
	\end{figure}
	The emergence of Hamiltonian chaos in regions typically occupied by plasma particles strongly suggests the breakdown of the adiabatic invariance of the magnetic moment, $\mu=\frac{\mathbf{v}_{\perp}^{2}}{2B}$. This quantity, which acts as a third integral of motion and would otherwise ensure system integrability, is no longer adiabatically invariant in such chaotic domains. Here, $\mathbf{v}_{\perp}$ denotes the velocity component perpendicular to the magnetic field, and $B$ is the magnetic field strength. To investigate this phenomenon, we adopt the same initial conditions as those depicted in \figref{-evolution-of-section} and track the evolution of the magnetic moment over time. To mitigate the rapid oscillations associated with cyclotron motion, we average $\mu$ over $500$ typical gyroperiods, and look at how far it deviates from its mean value (see \figref{Mu_mean}). The results, presented in \figref{-evolution-of-section}, compare two distinct trajectories: one corresponding to a \emph{quasiperiodic} orbit and the other to a \emph{chaotic} one. The quasiperiodic trajectory exhibits only minor --- almost but not quite --- periodic fluctuations in $\mu$, hinting at the preservation of the regularity of adiabatic motion. In stark contrast, the chaotic trajectory reveals a clear absence of adiabatic invariance, as evidenced by the erratic and unpredictable variations in the magnetic moment. These contrasting dynamics are further corroborated by their Fourier spectra, illustrated in \figref{Mu_mean}. The quasiperiodic trajectory yields a discrete spectrum, while the chaotic trajectory produces a broad, continuous spectrum. While the destruction of the adiabatic invariant has been documented for energetic particles in previous studies \cite{cambon_chaotic_2014,escande_breakdown_2021,Kamaletdinov2025}, its observation in thermal particles within a burning plasma represents, to our knowledge, a novel finding. \smallskip
	
	{\noindent\it Summary \& discussion.} In this letter, we have characterised classes of thermodynamic plasma equilibria derived from the principle of maximum entropy, in cylindrical geometry. These equilibria align with the well-established framework of magnetohydrodynamics (MHD) in the inviscid limit. Through perturbative analysis, these equilibria demonstrate robust dynamical stability when extended to toroidal configurations, as illustrated in \figref{-evolution-of-section}. Here, the trajectories of both a regular and a chaotic particle are depicted in physical space, revealing that chaotic motion does not necessarily entail large radial displacements. Instead, chaotic trajectories may remain spatially confined, even as they exhibit the hallmarks of Hamiltonian chaos. A striking observation is the destruction of the adiabatic invariance of the magnetic moment $\mu$, even for thermal particles in a burning plasma --- a phenomenon not previously documented. This breakdown raises critical questions: could these localised violations of $\mu$-invariance propagate across significant regions of phase space, ultimately leading to a global violation of adiabaticity? If so, the implications could undermine the foundational assumptions of gyrokinetic modelling in burning plasmas. 
	
	While the interplay between turbulence and energetic particles remains an open and complex problem, the present study highlights an additional layer of concern: chaos emerges even for thermal particles, which are both abundant and central to turbulent dynamics. This suggests the possibility of unaccounted transport losses, reduced flow generation, or enhanced flow damping --- mechanisms that could pose challenges distinct from those observed in current experimental devices. This phenomenon is specific to high-performance burning plasma scenarios and warrants further investigation, particularly in the presence of electrostatic or electromagnetic turbulence, in order to assess its broader impact on plasma confinement and stability. \\

\begin{acknowledgements}

This work has been carried out within the framework of the EUROfusion Consortium, funded by the European Union via the Euratom Research and Training Programme (Grant Agreement No 101052200 --- EUROfusion). Views and opinions expressed are however those of the author(s) only and do not necessarily reflect those of the European Union or the European Commission. Neither the European Union nor the European Commission can be held responsible for them.

\end{acknowledgements}

	\bibliographystyle{plain}

\end{document}